\documentclass[aps,prl,twocolumn]{revtex4} 
\usepackage{amsmath} 
\usepackage{graphicx} 
\usepackage{color}

\begin{document} 
\title{Twists in Ferromagnetic Monolayers With Trigonal Prismatic Symmetry} 

\author{Kjetil M.D.\ Hals$^1$ and Karin Everschor-Sitte$^2$} 
\affiliation{$^1$ Department of Engineering Sciences, University of Agder, 4879 Grimstad, Norway\\
$^2$ Institute of Physics, Johannes Gutenberg Universit\"at, 55128 Mainz, Germany} 
\newcommand{\Kjetil}[1]{\textcolor{red}{#1}} 
\newcommand{\Karin}[1]{\textcolor{green}{#1}}
\begin{abstract}
Two-dimensional materials such as graphene or hexagonal boron nitride are indispensable in industry. The recently discovered 2D ferromagnetic materials also promise to be vital for applications. In this work, we develop a phenomenological description of non-centrosymmetric 2D ferromagnets with trigonal prismatic crystal structure. We chose to study this special symmetry group since these materials do break inversion symmetry and therefore, in principle, allow for chiral spin structures such as magnetic helices and skyrmions. However, unlike all non-centrosymmetric magnets known so far, we show that the symmetry of magnetic trigonal prismatic monolayers neither allow for an internal relativistic Dzyaloshinskii-Moriya interaction (DMI) nor a reactive spin-orbit torque.  We demonstrate that the DMI only becomes important at the boundaries, where it modifies the boundary conditions of the magnetization and leads to a helical equilibrium state with a helical wavevector that is inherently linked to the internal spin orientation. Furthermore, we find that the helical wavevector can be electrically manipulated via dissipative spin-torque mechanisms.  
Our results reveal that 2D magnets offer a large potential for unexplored magnetic effects.
\end{abstract}

\maketitle 

The successful fabrication of isolated graphene in 2004~\cite{Novoselov:science2004} initiated an intense interest in manufacturing and exploring two-dimensional (2D) materials~\cite{Graphene:Review}. The last ten years\textsc{\char13} rapid progress within material engineering has revealed that graphene is just one of a whole family of stable 2D materials, which includes insulators, semiconductors as well as semimetals~\cite{2Dmaterials:Review1, 2Dmaterials:Review2}. In addition to graphene, other important examples are hexagonal boron nitride, transition metal dichalcogenides (TMDs), and phosphorene~\cite{2Dmaterials:Review1}. Recent experiments have demonstrated that the TMDs even become superconducting at low temperatures~\cite{Wang:np2016}. The unique tunable electronic and optical properties of the 2D materials, combined with their mechanical flexibility and stretchability, make them particularly attractive as building blocks to create novel quantum materials with the potential for integration in the next generation of electronic devices. 

A novel member of this family is the 2D ferromagnets. In 2017, two independent research groups experimentally demonstrated long-range ferromagnetic ordering in monolayers and bilayers of CrI$_3$ and Cr$_2$Ge$_2$Te$_6$ at temperatures below 61~K and 30~K respectively~\cite{Huang:nature2017, Gong:nature2017}. The experiments showed only a weak coupling of the spins to the adjacent substrates, thus indicating that the ordered spins could be regarded as purely 2D ferromagnetic systems. Furthermore, a recent experiment reported strong ferromagnetic ordering in monolayers of the TMD VSe$_2$ at temperatures far above 300~K~\cite{Bonilla:nn2018} . The observed room-temperature ferromagnetism in TMDs makes this class of materials particularly promising for spintronics applications. In particular, it is believed that their remarkable 2D properties will unlock a multitude of new exotic spin phenomena with potential applications in the development of novel ultra-compact spin-based devices.   

\begin{figure}[ht] 
\centering 
\includegraphics[scale=1.0]{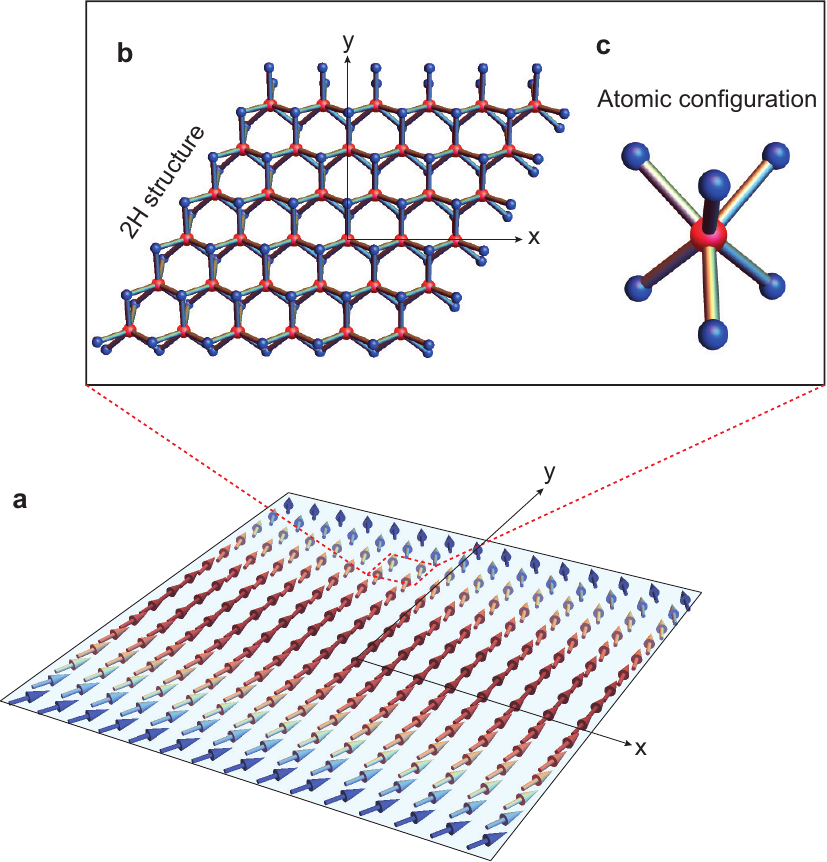}  
\caption{(color online). (a) Ferromagnets with 2H structure lack inversion symmetry. Nevertheless, the systems do not have an internal DMI field.  Only a boundary-DMI field exists, which in equilibrium stabilizes a helical spin phase (arrows).
(b) The crystal structure of the 2H monolayer. 
(c) Shows the atomic configuration at a lattice point. In the TMDs, the red atom represents the transition metal atom and the blue atoms are the chalcogen atoms. }
\label{Fig1} 
\end{figure} 

The TMDs crystallize in two different phases, either in the 1T-phase or in the 2H-phase~\cite{2Dmaterials:Review1}. The 1T-phase is characterized by the octahedral point group $D_{3d}$, whereas the 2H-phase is determined by the trigonal prismatic point group $D_{3h}$ (see Fig.~\ref{Fig1}). The experiment on the magnetic monolayer in Ref.~\cite{Bonilla:nn2018} was performed for VSe$_2$ in the centrosymmetric 1T-phase. Ferromagnetism in the 2H-phase has not yet been experimentally observed. 
However, theory predicts that nano sheets of ferromagnetic TaS$_2$ exist in the 2H-phase~\cite{2Dmaterials:Review1}.

We anticipate the spatially asymmetric 2H-phase to exhibit more intriguing spin physics than the centrosymmetric 1T-phase. Generally, broken inversion symmetry combined with spin-orbit coupling (SOC) yields an anisotropic relativistic exchange interaction known as the Dzyaloshinskii-Moriya interaction (DMI)~\cite{Dzyaloshinsky:jpcs1958, Moriya:pr1960}. Several experiments have demonstrated that the DMI gives rise to complex physical effects such as helical spin phases~\cite{Dzyaloshinsky:jpcs1958}, helimagnons~\cite{Moller:prl1967, Janoschek:prb2010}, boundary-induced twist states~\cite{Rohart:prb2013, Meynell:prb14, Rybakov:prb13, Muller:njp16, Leonov:prl16, Hals:prl2017, Meynell:prb17, Leonov:natcom17, Mulkers:prb2018}, 
and topologically stabilized magnetic skyrmion textures~\cite{Nagaosa:natnano13, Wiesendanger:natrevmat16, Kang:ieee16, Finocchio:jphysd16,Leonov:njp16, Garst:jphysd17, Jiang:phsrep17, Fert:natrevmat17}.  
Additionally, spatial asymmetric SOC strongly affects the coupling between the magnetization and the itinerant charge carriers and leads to novel current-driven spin-torque mechanisms such as the spin-orbit torques (SOTs) and spin-Hall torque~\cite{Brataas:nn14}.  

So far, there have been no previous studies of the SOC-induced spin phenomena in ferromagnetic monolayers with 2H structure.
Importantly, the relativistic effects can be particularly strong in the trigonal prismatic state, because first-principle calculations indicate that the phase is often semiconducting~\cite{2Dmaterials:Review2, Bonilla:nn2018}. Consequently, we expect the SOC and the potentially accompanying DMI  and SOTs to be particularly strong. 

Here, we develop a phenomenology of the magnetization dynamics in 2D ferromagnets with 2H structure. In stark contrast to previously investigated ferromagnets with broken inversion symmetry, we find that the 2H-ferromagnets exhibit \emph{no internal DMI field and no reactive SOT}. This is a direct consequence of the trigonal prismatic crystal symmetry of the 2H phase. 
Remarkably, the DMI reveals itself only close to the edges of the sample, where the asymmetric exchange interaction, denoted as boundary-DMI in the following, modifies the micromagnetic boundary conditions (BCs).  We demonstrate that the DMI-induced BCs produce a helical modulation of the magnetic moments with a helical wavevector that is locked to the internal orientation of the magnetization. Importantly, we show that this unique locking enables all-electric manipulation of the helical wavevector via the dissipative SOT and spin-transfer torque (STT).

To derive a phenomenological description of 
ferromagnetic materials with trigonal prismatic 2H structure, we start by writing down the magnetic free-energy functional dictated by the $D_{3h}$ symmetry. 
We parameterize the local magnetization direction by the unit vector $\mathbf{m} (\mathbf{r},t)$ and expand the free energy up to second order in the magnetization gradients. In this case, the free energy
can be expressed as~\cite{LL:book}
\begin{equation}
F[ \mathbf{m} ]  = \int {\rm d\mathbf{r}}\left[  \mathcal{F}_{\rm e} + \mathcal{F}_{\rm D} + \mathcal{F}_{\rm h} + \mathcal{F}_{\rm a} \right], \label{Eq:F1}
\end{equation}
where  $\mathcal{F}_{\rm e}=J_{ij}\,  \partial_i \mathbf{m}\cdot \partial_j \mathbf{m}$ represents the symmetric exchange interaction, 
$\mathcal{F}_{\rm D}= D_{ijk} m_i \partial_j m_k$ is the general form of DMI \cite{LL:book}, 
 $\mathcal{F}_{\rm h}= -\mathbf{m}\cdot \mathbf{H}_{\rm ext}$ describes the coupling to an external magnetic field, 
$\mathcal{F}_{\rm a}= K_{ij}m_im_j$ represents the anisotropy energy,
and the integration is over the 2D plane covered by the monolayer.

The tensors $J_{ij}$, $K_{ij}$, and $D_{ijk}$ are polar tensors of rank two and three respectively, and their tensorial structures are governed by the symmetry of the system. For systems characterized by the $D_{3h}$ point group, polar second-rank tensors have three non-vanishing tensor elements that are parameterized by two independent parameters~\cite{Birss:book}. 
In the following, we choose a coordinate system where the $z$-axis is parallel to the three-fold rotation axis and $y$ is along a two-fold rotation axis, see Fig. 1.
In this reference frame, the ferromagnetic monolayer lies in the $xy$-plane and the symmetric exchange interaction and anisotropy energy have the following non-zero coefficients: 
 $J_{xx}=J_{yy}\equiv J $,  $J_{zz}$ and $K_{xx}=K_{yy}$, $K_{zz}$.  
The DMI tensor has four non-vanishing tensor elements that are parameterized by a single parameter $D$~\cite{Birss:book}:
\begin{equation}
D\equiv D_{xyy}= D_{yxy}= D_{yyx}= - D_{xxx}.\label{Eq:Dijk}
\end{equation}
Substitution of the above tensor coefficients into Eq.~\eqref{Eq:F1} yields the free energy density   
\begin{eqnarray}
\mathcal{F} &=&  J\left[ \left( \partial_x\mathbf{m}\right)^2 + \left(\partial_y\mathbf{m}\right)^2 \right]  - \mathbf{m}\cdot\mathbf{H}_{\rm ext}   + K_u m_z^2 + \label{Eq:F2} \\
& & D\left[  m_y\partial_ym_x + m_y\partial_xm_y + m_x\partial_ym_y  - m_x\partial_x m_x   \right]  . \nonumber
\end{eqnarray}
Here, we have introduced the uniaxial anisotropy constant $K_u = K_{zz} - K_{xx}$ and only taken into account spatial variations in the $xy$-plane.

The magnetization dynamics are given by  the Landau-Lifshitz-Gilbert (LLG) equation~\cite{Ralph:jmmm08,Brataas:nature2012}
\begin{equation}
\dot{\mathbf{m}}  = -\gamma\mathbf{m}\times\mathbf{H}_{\rm eff} + \mathbf{m}\times\boldsymbol{\alpha}\dot{\mathbf{m}} + \boldsymbol{\tau}, \label{Eq:LLG}
\end{equation} 
where $\gamma$ is the gyromagnetic ratio, $\mathbf{H}_{\rm eff}$ is the effective field and $\boldsymbol{\tau}$ the current-induced torque. 
The matrix $\boldsymbol{\alpha}$ is the Gilbert damping tensor, which is a second-rank polar tensor with the non-vanishing elements
$\alpha_{xx}=\alpha_{yy}= \alpha_{\bot}$ and $\alpha_{zz}$.
The effective field $\mathbf{H}_{\rm eff}$ is determined by the internal value of the functional derivative,
\begin{equation}
\mathbf{H}_{\rm eff} = -\frac{\delta \mathcal{F}}{ \delta \mathbf{m}}=2 J  \left[  \partial_x^2\mathbf{m}  + \partial_y^2\mathbf{m}  \right] + \mathbf{H}_{\rm ext} - 2K_u m_z\hat{\mathbf{z}}, \label{Eq:Heff}
\end{equation}  
whereas the boundary terms in the variational equation $\delta F / \delta \mathbf{m}= 0$ lead to the following micromagnetic BCs of the LLG equation~\cite{Hals:prl2017}: 
\begin{equation}
2J\hat{\mathbf{n}}\cdot\boldsymbol{\nabla}\mathbf{m} = -\mathbf{m}\times\left( \boldsymbol{\Gamma}_D \times\mathbf{m} \right) . \label{Eq:BC}
\end{equation}
Here, $\hat{\mathbf{n}}= [n_x, n_y]^T$ ($T$ denotes the transpose of the vector) is the surface normal and $\boldsymbol{\Gamma}_D$ is the boundary-induced DMI field:
\begin{equation}
\boldsymbol{\Gamma}_D = D \left[ m_yn_y - m_xn_x, m_xn_y + m_yn_x, 0 \right]^T. \label{Eq:BoundaryField}
\end{equation}

Eqs.~\eqref{Eq:LLG}-\eqref{Eq:BoundaryField} are the first central result of this work and represent a general phenomenological theory of the magnetization dynamics in 2H-ferromagnets.
Importantly, we find that the DMI \emph{only} affects the BCs in such systems. Despite the fact that 2H-ferromagnets have broken spatial inversion symmetry, the DMI does not enter the effective field $\mathbf{H}_{\rm eff}$  in Eq.~\eqref{Eq:Heff}. The reason for this is that only the parts of the DMI tensor, which are antisymmetric with respect to the magnetic indices
$D_{[ik]j}= D_{ijk} - D_{kji}$~\cite{Hals:prl2017},  appear in the effective field $(H_D)_k = (D_{ijk}-D_{kji}) \partial_j m_i$. These, however, vanish for systems with trigonal prismatic crystal symmetry, see Eq.~\eqref{Eq:Dijk}. 
This differs markedly from other non-centrosymmetric magnets, in which the main effect of the DMI
is to produce an internal effective field that favors a helical modulation of the magnetization direction.
With having derived Eq.~\eqref{Eq:BC}, we predict that despite the absence of internal DMI, the DMI will still influence the magnetization at the boundaries via the Neumann BCs.

Next, we will use the above formalism to investigate the equilibrium state of ferromagnetic 2H monolayers.
We assume a strong in-plane alignment of the spins ($K_u \gg 0$), which is consistent with the recent experiment on VSe$_2$~\cite{Bonilla:nn2018}. 
In this case, the magnetic state is completely determined by the azimuthal angle $\phi (\mathbf{r}, t)$:
\begin{equation}
\mathbf{m} (\mathbf{r},t)= \left[ \cos (\phi), \sin (\phi), 0  \right]^T . \label{Eq:m}
\end{equation}
For simplicity, we disregard external magnetic fields.

The equilibrium equations for the magnetization are found from the static LLG equation, i.e., when $\dot{\mathbf{m}}=0$ in Eq.~\eqref{Eq:LLG}. Substituting Eq.~\eqref{Eq:m} into Eqs.~\eqref{Eq:LLG}-\eqref{Eq:BoundaryField}  produces the following 
 boundary-value-problem (BVP) for $\phi$:
\begin{align}
\tilde{\nabla}^2 \phi &= 0, \ \  & \text{internally}, \label{Eq:BVP1} \\
\hat{\mathbf{n}}\cdot \tilde{\boldsymbol{\nabla}} \phi &= -\tilde{D}\hat{\mathbf{n}}\cdot \hat{\mathbf{f}} \left( \phi \right), \ \  & \text{at boundaries}. \label{Eq:BVP2}
\end{align}

Here, we have scaled the axes by a length scale $a$ that characterizes the typical size of the sample, $\tilde{x}= x/a$ and $\tilde{y}= y/a$, and introduced the dimensionless 2D nabla operator     
$\tilde{\boldsymbol{\nabla}}= \left[ \partial_{\tilde{x}}, \partial_{\tilde{y}} \right]^T$ and DMI parameter $\tilde{D}= D a/2J$. 
The unit vector $\hat{\mathbf{f}} $ is
a function of $\phi$ and given by $ \hat{\mathbf{f}} \left( \phi \right)^T = \left[ \sin (2\phi), \cos (2\phi) \right]= [2m_xm_y, m_x^2-m_y^2]$.
We see that $\phi$ is governed by the 2D Laplace equation with nonlinear inhomogeneous Neumann BCs.
Qualitatively, we expect for larger discs that the internal equation plays the dominant role, while the BCs become more important for smaller systems. We note, however, that for very small samples our continuum description is not applicable.

The effect of the DMI-induced BCs is to produce a magnetic twist state at the edges of the sample. Such boundary-induced twist states have been studied in ferromagnetic heterostructures, 
where a DMI field occurs in the bulk of the sample. In contrast, the 2H-ferromagnets are only affected by the DMI via the BCs in Eq.~\eqref{Eq:BVP2}. 

Note that $\hat{\mathbf{n}} $ and $\hat{\mathbf{f}} $ consist of respectively the linear and quadratic basis functions for the irreducible representation $E^{'}$ of $D_{3h}$. 
Consequently, the BCs \eqref{Eq:BVP2} are as expected invariant under any symmetry transformation of $D_{3h}$ that acts simultaneously on $\hat{\mathbf{n}} $ and $\hat{\mathbf{f}} $.

While being linear in the relativistic interactions, the DMI is much smaller than the exchange interaction. Therefore, we expect $\tilde{D}$ to be a small parameter in our problem, which allows us to solve the BVP in Eqs.~\eqref{Eq:BVP1}-\eqref{Eq:BVP2} perturbatively to first order in $\tilde{D}$. 
To this end, we consider a perturbative solution for $\phi$ of the form
\begin{equation}
\phi (\tilde{\mathbf{r}}) = \phi_0 + \tilde{D}\phi_1 (\tilde{\mathbf{r}}). \label{Eq:PhiExp}
\end{equation}
The zeroth-order solution $\phi_0$ represents the magnetization direction in absence of any DMI, while  
the first-order solution $\phi_1$ determines the boundary-driven spatial modulation of the magnetization direction. 

Further, we will assume that our sample has the shape of a disk with a rescaled radius of $R/a= 1$. 
Upon substitution of Eq.~\eqref{Eq:PhiExp} into Eqs.~\eqref{Eq:BVP1}-\eqref{Eq:BVP2}, we find the equations for $\phi_0$ and $\phi_1$. The zeroth-order contribution is given by the 2D Laplace equation
$\tilde{\nabla}^2\phi_0=0$ with the homogeneous Neumann BCs $\hat{\mathbf{n}}\cdot \tilde{\boldsymbol{\nabla}} \phi_0 = 0$. The solution of this equation is just a constant that represents the internal value of the magnetization direction. 
The first-order correction to this solution is given by the BVP
\begin{align}
\tilde{\nabla}^2 \phi_1 &= 0, \ \  &  \text {when } \tilde{x}^2 + \tilde{y}^2 < 1,  \label{Eq:BVP1b} \\
\hat{\mathbf{n}}\cdot \tilde{\boldsymbol{\nabla}} \phi_1 &= -\hat{\mathbf{n}}\cdot \hat{\mathbf{f}} \left( \phi_0 \right), \ \  & \text {when } \tilde{x}^2 + \tilde{y}^2 = 1. \label{Eq:BVP2b}
\end{align} 
Via separation of variables, one obtains $\phi_1= -\tilde{x} \sin (2\phi_0) - \tilde{y} \cos (2\phi_0)$.
Thus, to first order in the relativistic interactions, we find the solution
\begin{equation}
\phi (\tilde{\mathbf{r}} ) = \phi_0  + \tilde{\mathbf{k}}_D\cdot \tilde{\mathbf{r}}, \label{Eq:phi}
\end{equation}
where $\tilde{\mathbf{r}} = (\tilde{x}, \tilde{y})^T$ is the position vector on the unit disk. 
The dimensionless wavevector $\tilde{\mathbf{k}}_D$ is determined by
\begin{equation}  
\tilde{\mathbf{k}}_D = - \tilde{D} [\sin (2\phi_0), \cos (2\phi_0)]^T.   \label{Eq:kD}
\end{equation}

\begin{figure}[ht] 
\centering 
\includegraphics[scale=1.0]{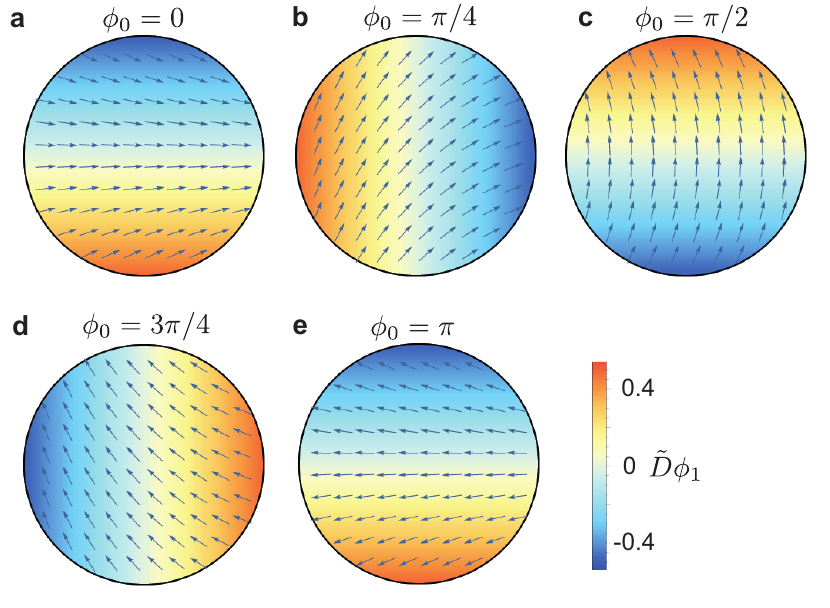}  
\caption{(color online). (a)-(e) Shows the equilibrium state of the magnetization (arrows) on the 2D disk for $\phi_0=0$, $\phi_0=\pi / 4$, $\phi_0=\pi / 2$, $\phi_0=3\pi / 4$, and $\phi_0=\pi $, respectively.
The density plots illustrate the first-order correction $\tilde{D}\phi_1 $ to the magnetic equilibrium state produced by the DMI-induced BCs. For clarity, we have used $\tilde{D}= 0.5$.}
\label{Fig2} 
\end{figure} 

Eqs.~\eqref{Eq:phi}-\eqref{Eq:kD} represent the second central result of this work and demonstrate that the DMI-induce Neumann BCs \eqref{Eq:BVP2} to first order in $\tilde{D}$ produces
a helical modulation of the magnetization direction. Significantly, we find that the direction of the helical  wavevector $\tilde{\mathbf{k}}_D$ is inherently
linked to the internal orientation of the magnetization. 
The functional relationship $\tilde{k}_{D,x}\sim \sin (2\phi_0)$ ($\tilde{k}_{D,y}\sim \cos (2\phi_0)$) between $\tilde{\mathbf{k}}_D$ and the magnetization implies that the wavevector has a two-fold symmetry with respect to $\phi_0$.  
This is illustrated in Fig.~\ref{Fig2}, which shows the function $\tilde{D} \phi_1$  and the magnetization in Eq.~\eqref{Eq:m} for five different values of $\phi_0$. 

The locking of $\tilde{\mathbf{k}}_D$ to the internal magnetization opens the possibility to electrically control the helical spin phase via current-driven magnetization torques. 
In what follows, we study the effects of the STT and SOT to first order in the applied current-density $\boldsymbol{\mathcal{J}}$,  the magnetization gradients and the degree of magnetic anisotropy, as well as to lowest order in the relativistic interactions. In this case, the STT takes the form of $\boldsymbol{\tau}_{\mathrm{STT}}= (1-\beta\mathbf{m}\times)(P\boldsymbol{\mathcal{J}}\cdot\boldsymbol{\nabla})\mathbf{m}$, where $P$ is proportional to the spin polarization of the current and $\beta$ is the non-adiabatic torque parameter~\cite{Hals:prb13,Hals:prb15}. The SOT is $\boldsymbol{\tau}_{\mathrm{SOT}}= -\gamma \mathbf{m}\times \mathbf{H}_{\mathrm{SOT}}$,
with the current-induced effective field  $H_{\mathrm{SOT},i}= (\Lambda_{ij}^{(r)} + \Lambda_{ijk}^{(d)}m_k) \mathcal{J}_j$. 
 $\Lambda_{ij}^{(r)}$ ($\Lambda_{ijk}^{(d)}$) represents the reactive (dissipative) contribution to the SOT~\cite{Hals:prb13,Hals:prb15}.

The tensors $\Lambda_{ij}^{(r)}$ and $\Lambda_{ijk}^{(d)}$ originate from the intrinsic SOC and their tensorial structures are solely determined by the crystallographic point group of the system~\cite{Hals:prb13,Hals:prb15}.
Interestingly, the $D_{3h}$ group implies that $\Lambda_{ij}^{(r)} = 0$~\cite{Birss:book}. As a result, \emph{ferromagnetic trigonal prismatic monolayers do not exhibit any reactive SOT} under the application of an external electric field.
This is in stark contrast to previously studied asymmetric ferromagnets, in which the reactive SOT is the dominant current-driven torque mechanism (see Ref.~\cite{Brataas:nn14} and references therein). 
Only the dissipative SOT is present in ferromagnets with $D_{3h}$ symmetry and its tensor is determined by a single parameter $\Lambda_{\mathrm{so}}$. The non-vanishing tensor elements are~\cite{Birss:book}
\begin{equation}
 \Lambda_{\mathrm{so}} \equiv \Lambda_{yyx}^{(d)} = \Lambda_{yxy}^{(d)} = \Lambda_{xyy}^{(d)} = -\Lambda_{xxx}^{(d)},
 \end{equation}
in analog to Eq.~\eqref{Eq:Dijk}.
As the dissipative SOT is governed by the momentum space Berry curvature~\cite{Kurebayashi:nn14}, the strength of the SOT in ferromagnetic trigonal prismatic monolayers is strongly linked to
the band structure topology of the system. In particular, the appearance of band crossings close to the Fermi surface will greatly enhance the strength of the Berry curvature and thus the SOT.

\begin{figure}[ht] 
\centering 
\includegraphics[scale=1.0]{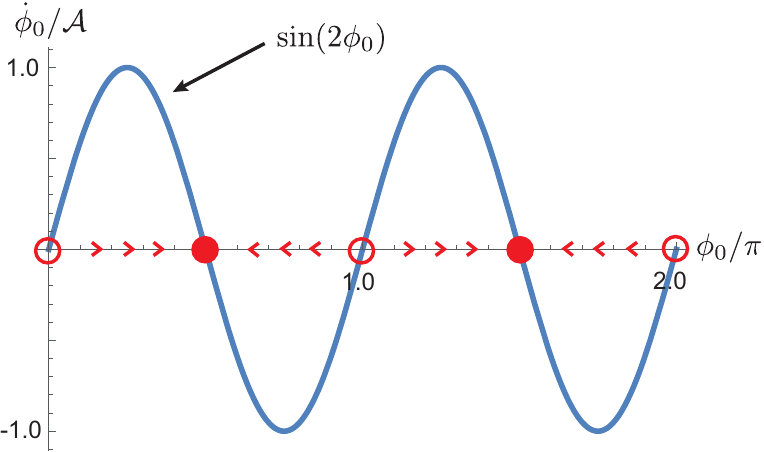}  
\caption{(color online). Vectorfield plot of the differential equation in Eq.~\eqref{Eq:DotPhi} with the current density applied along the $x$-axis. 
The red circles indicate unstable fix points for the solution $\phi_0 (t)$, whereas the red dots represent stable fix points. 
Here, we assume $\mathcal{A}\equiv -\frac{|\boldsymbol{\mathcal{J}}|\tilde{D}}{\alpha_{\bot}} \left( \frac{P\beta}{R} - \frac{\gamma\Lambda_{so}}{\tilde{D}}  \right) > 0$. }
\label{Fig3} 
\end{figure}

For modeling the current-induced magnetization dynamics, we apply a collective coordinate description assuming that  the magnetic state is described by the perturbative solution \eqref{Eq:phi}. 
The Thiele equation for the time dependent collective angle $\phi_0 (t)$ is given by $\Gamma \dot{\phi}_0 = L$, where 
$L= \int {\rm d}\mathbf{r} \, \partial \mathbf{m}/ \partial\phi_0 \cdot (\mathbf{m}\times (\boldsymbol{\tau}_\mathrm{STT}+\boldsymbol{\tau}_\mathrm{SOT}))$ and 
$\Gamma= \int {\rm d}\mathbf{r} \alpha_{\bot} (\partial \mathbf{m}/ \partial\phi_0)^2$.
The resulting equation becomes
\begin{equation}
\dot{\phi}_0 = \frac{1}{\alpha_{\bot}} \left( \frac{P\beta}{R} - \frac{\gamma\Lambda_{so}}{\tilde{D}}  \right)  \boldsymbol{\mathcal{J}} \cdot \tilde{\mathbf{k}}_D.\label{Eq:DotPhi}
\end{equation}  
Both the dissipative STT and SOT contribute to the dynamics.
Note that the current density couples directly to $\mathbf{k}_D \sim \sin(2\phi_0)$ and thus directly to the orientation of the helix. 
For any applied current, the two-fold symmetry of $\mathbf{k}_D$ induces two (un-)stable fix points, i.e.\ (un-)preferred directions for $\phi_0$ (see Fig.~\ref{Fig3}). 
As the locations of these fix points can be tuned by changing the direction of the current, the helix can be orientated along any axis in the $xy$-plane via the electric field.  
The current-driven effects on the boundary-induced helix are very different from those on bulk helical textures, where an applied current usually only leads to a weak tilting of the bulk magnetization~\cite{Footnote}.
The boundary-induced spin texture therefore opens the door for a unique way of controlling the helical wavevector solely by means of electric currents.

In summary, we have developed a phenomenological description of the magnetization dynamics in 2D ferromagnetic materials with trigonal prismatic  2H structure.  Extraordinarily, we find that the system exhibits no internal DMI field, despite the fact that the 2H structure lacks inversion symmetry. We derive that the DMI only affects the spin physics at the boundaries, where it yields non-trivial BCs for the magnetization. The DMI-induced BCs turn a field-polarized phase into a helical phase, where the direction of the helical wavevector is locked to the ferromagnetic orientation. By symmetry considerations we show that these systems are subject to dissipative STTs and SOTs, which allow to electrically control the direction of the helical wavevector.


\end{document}